\documentclass[ChapterTOCs]{EEH}
\usepackage{amssymb}
\usepackage{amsmath}
\usepackage{graphicx}
\usepackage{subfigure}
\usepackage[sectionbib]{bibunits}
\usepackage{makeidx}
\usepackage{multicol}
\usepackage{url}
\makeindex

\begin{document}

\bibliographyunit[\chapter]
\defaultbibliography{chapterMI}
\defaultbibliographystyle{alpha}
\setcounter{page}{1}

\chapter{Comb models for transport along spiny dendrites }

\begin{chapterauthors}
\chapterauthor{V.  M\'endez}{Universitat Aut\`onoma de Barcelona. Spain}
\chapterauthor{A. Iomin}{Technion - Israel Institute of Technology, Israel}
\end{chapterauthors}

\section{Introduction}
\begin{figure}
\centerline{\includegraphics[scale=0.8]{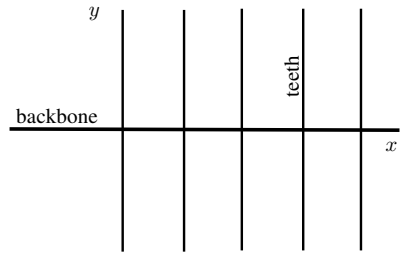}}
\caption{Two sided comb}\label{fig:1}
\end{figure}

A comb is a simplified model for various types of natural
phenomena which belong to the loopless graphs category. The comb
consists of a backbone along the horizontal axis and fingers or
teeth along the perpendicular direction (see Figure \ref{fig:1}
for a two sided comb).

\begin{figure}
\centerline{\includegraphics[scale=0.6]{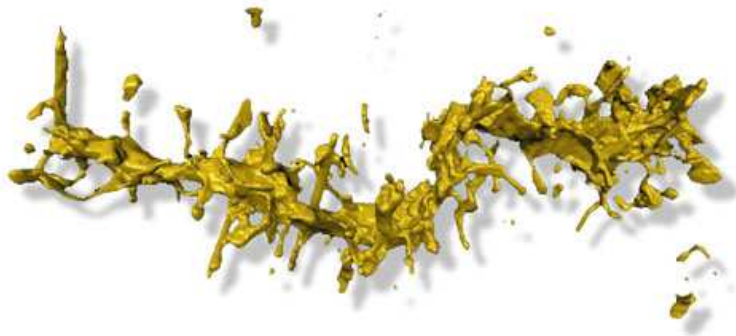}}
\caption{Electron tomogram of a spiny dendrite. Image taken from
Internet
(http://www.cacr.caltech.edu/projects/ldviz/results/levelsets/).}
\label{fig:1bis}
\end{figure}

Comb like models have been applied to mimic ramified structures as
spiny dendrites of neuron cells \cite{MeIo13, IoMe13} or
percolation clusters with dangling bonds \cite{WeHa86}. We are
interested in the first example, where a comb structure with one
sided teeth of infinite length can be used to describe the
movement and binding dynamics of particles inside the spines of
dendrites. These spines are small protrusions from many types of
neurons located on the surface of a neuronal dendrite. They
receive most of the excitatory inputs and their physiological role
is still unclear although most spines are thought to be key
elements in neuronal information processing and plasticity
\cite{Yu10}. Spines are composed of a head ($\sim 1$ $\mu$m) and a
thin neck ($\sim 0.1$ $\mu$m) attached to the surface of dendrite
(see Fig. \ref{fig:1bis}).

The heads of spines have an active membrane, and as a consequence,
they can sustain the propagation of an action potential with a
rate that depends on the spatial density of spines \cite{EaBr09}.
Decreased spine density can result in cognitive disorders, such as
autism, mental retardation and fragile X syndrome \cite{nim}.
Diffusion over branched smooth dendritic trees is basically
determined by classical diffusion and the mean square displacement
(MSD) along the dendritic axis grows linearly with time. However,
inert particles diffusing along dendrites enter spines and remain
there, trapped inside the spine head and then escape through a
narrow neck to continue their diffusion along the dendritic axis.
Recent experiments together with numerical simulations have shown
that the transport of inert particles along spiny dendrites of
Purkinje and Pyramidal cells is anomalous with an anomalous
exponent that depends on the density of spines \cite{SaWi06,
SaWi11, ZeHo11}. Based on these results, a fractional
Nernst-Planck equation and fractional cable equation have been
proposed for electrodiffusion of ions in spiny dendrites
\cite{HeLaWe08}. Whereas many studies have been focused to the
coupling between spines and dendrites, they are either
phenomenological cable theories \cite{HeLaWe08,phen} or
microscopic models for a single spine and parent dendrite
\cite{mic,mic2}. More recently a mesoscopic non-Markovian model
for spines-dendrite interaction and an extension including
reactions in spines and variable residence time have been
developed \cite{prl,MeFeHo10}. These models predict anomalous
diffusion along the dendrite in agreement with the experiments but
are not  able to relate how the anomalous exponent depends on the
density of spines \cite{SaWi11,ZeHo11}. Since these experiments
have been performed with inert particles (i.e., there are not
reaction inside spines or dendrites) one concludes that the
observed anomalous diffusion is due exclusively to the geometric
structure of the spiny dendrite. Recent studies on the transport
of particles inside spiny dendrites indicate the strong relation
between the geometrical structure and anomalous transport
exponents \cite{SaWi11,bbz,kubota}. Therefore, elaboration such an
analytic model that establishes this relation can be helpful for
further understanding transport properties in spiny dendrites. The
real distribution of spines along the dendrite, their size and
shapes are completely random \cite{nim}, and inside spines
the spine necks act as a transport barrier \cite{mic}. For these
reasons one may reasonably assume that the diffusion inside spine
is anomalous. In this chapter, we describe some models,
based on comb-like structure that mimic a spiny dendrite; where the
backbone is the dendrite and the teeth (lateral branches) are the
spines. The models predict: i) anomalous transport inside spiny
dendrites, in agreement with the experimental results of Ref.
\cite{SaWi06}, ii) also explain the dependence between the mean
square displacement and the density of spines observed in
\cite{SaWi11} and iii) the mechanism of translocation wave of
CaMKII (${\rm Ca}^{2+}$ - calmodulin-dependent protein kinase II,
a key regulator of the synaptic function). The chapter is
organized as follows. First we study the statistical properties of
combs and explain how to reduce the effect of teeth on the movement along
the backbone as a waiting time  distribution between
consecutive jumps.
Second, we justify an employment of a comb-like structure
as a paradigm for further exploration of a spiny dendrite. In particular,
we show how a comb-like structure can sustain the phenomenon of
the anomalous diffusion, reaction-diffusion and L\'evy walks.
Finally, we illustrate how the same models can be also useful to
deal with the mechanism of ta translocation wave / translocation waves of CaMKII and its
propagation failure.  We also present a brief introduction to the fractional integro-differentiation in appendix at the end of the chapter.

\section{Random walks in combs}
The statistical properties of comb-like structures have been
widely studied in the last century. The first passage-time and the
survival probability were studied by some authors
\cite{WeHa86,Re01}. More recently the interest has been centered
in the waiting time distribution equivalent to perform a random
walk along the teeth \cite{VdB89,CaMe05,Da07,MeFeHo10}, the mean
encounter time between two random walkers \cite{Ag14} and the
occupation time statistics \cite{ReBa13}. In this section we
illustrate two methods to compute the waiting time distribution
that mimics the effect of a random walk along a teeth.

We first consider the case of a discrete 1D chain where the
nearest neighbors are separated by a distance $a$. A random walk,
where each walker moves only to one of its nearest neighbors with
equal probability after a fixed waiting time $\tau$, is
characterized by the following probability distribution functions (pdf)s for the waiting times and
jump lengths, respectively,

\begin{eqnarray}
\phi(t) &=&\delta (t-\tau ),   \nonumber \\
w(x) &=&\frac{1}{2}\left[ \delta (x-a)+\delta (x+a)\right].
\label{eq:c5s4e1}
\end{eqnarray}

\begin{figure}
\centerline{\includegraphics[scale=0.8]{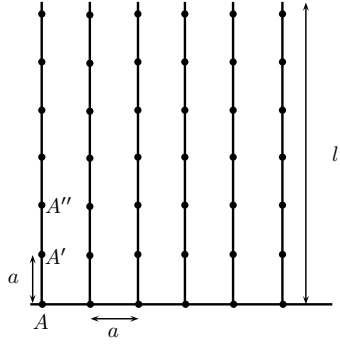}}
\caption{Sketch of a comb structure with a
nearest-neighbor distance $a$ and teeth of length $l$. The symbols
$A$, $A'$, $A''$... denote sites on the lattice
(see text).}
\label{fig:comb}
\end{figure}

In this way, systems with discrete time and space can be analyzed
in terms of the Continuous-Time-Random -Walk (CTRW). Next, we add
to every site of the backbone a secondary branch of length $l$, to
produce a one sided comb-like structure (see Fig.~\ref{fig:comb}).
On such a structure, a walker that is at a given site of the
backbone, can spend a certain amount of time in the secondary
branch before jumping to one of the nearest neighbor sites on the
backbone. If we are only interested in the behavior of the system
in the direction of the backbone, then the secondary branches
introduce a delay time for jumps between the neighboring sites on
the backbone. The random walk on the comb structure can be
modelled as a CTRW with (\ref{eq:c5s4e1}) and a renormalized
waiting pdf $\phi (t)$ that includes the effect of the delay due
the motion along teeth.

To determine analytically the effect of the teeth, we invoke
convolution rules that were introduced in \cite{VdB89} for the
case of homogeneous lattices:

(i) Consider a walker that is initially at a certain site within a
tooth. If the walker proceeds further into a tooth, i.e., moves
away from the backbone, its probability to return to the initial
site after a time $t$ is a convolution of factors, i.e., a product
in the Laplace space.

(ii) The total probability for that walker to return to the
initial site is determined by summing over all $t$ from $0$ to
$\infty$.

(iii) When the walker reaches a crossing, where it can choose
between different directions, the total probability is the sum of
the probabilities for each possible direction.

The teeth have no other additional crossings. It is as
shown in Fig.~\ref{fig:comb}. For the sake of generality, we
consider the case that when the walker is at a site on the
backbone, it can jump to another site on the backbone with
probability $\alpha$, or move onto the secondary branch with
probability $1-\alpha $.

Without loss of generality, we assume that initially the walker is
located on the backbone, and we apply the rules (i) -- (iii) to
determine $\phi(t)$. We study three specific cases.

(a) \textit{Comb structure with} $l=a$. In this case there is only
one site on the tooth, $A^{\prime}$. The walker can only jump in
the direction of the backbone with probability $\alpha$ or move
onto the branch with probability $1-\alpha$ and then return to the
initial site at the next jump. The time it takes to reach one of
the nearest neighbors on the backbone is $t=\tau$ with probability
$\alpha$, $t=3\tau$ with probability $(1-\alpha )\times 1\times
\alpha$, $t=5\tau$ with probability $(1-\alpha )^{2}\times
1^{2}\times \alpha$, and so on. We can write intuitively the
general form $\phi(t)$ as
\begin{equation}
\phi (t)=\sum_{j=1}^{\infty }\alpha (1-\alpha )^{j-1}\delta \left[%
t-\left(2j-1\right) \tau \right] .
\label{eq:c5s4e5}
\end{equation}

The rules listed above for $\phi(t)$ should reproduce this
behavior. For this purpose we need to work in the Laplace space.
Let $\hat{\phi}(s)$ be the Laplace transform of $\phi(t)$. The
rules (i) -- (iii) lead to the expression
\begin{equation}
\hat{\phi}(s)=\alpha \hat{\phi}_{0}\sum_{j=0}^{\infty}\left[(1-\alpha )%
\hat{\phi}_{0}^{2}\right]^{j}=\frac{\alpha\hat{\phi}_{0}}{1-(1-\alpha )%
\hat{\phi}_{0}^{2}}\, , \label{eq:c5s4e6}
\end{equation}%
where $\hat{\phi}_{0}$ is the probability distribution for a single
jump, $\hat{\phi}_{0}=e^{-\tau s}$, which is the Laplace transform of
(\ref{eq:c5s4e1}).

Equation (\ref{eq:c5s4e6}) is derived as follows. The term
$(1-\alpha)\hat{\phi}_{0}^{2}$ in the sum represents, according to
rule (i), the probability function for each occurrence of the
walker moving onto the secondary branch. This expression must be
summed up to infinity, according rule (ii), to take into
account that the walker can move onto the tooth
$1,2,...,\infty$ times. The factor $\alpha\hat{\phi}_{0}$
accounts for the final jump to the nearest neighbor on the
backbone.

It is easy to see that the expression (\ref{eq:c5s4e6}) may be
written as a Taylor series,
\begin{equation}
\hat{\phi}(s)=\sum_{j=1}^{\infty}\alpha (1-\alpha )^{j-1}\left(%
\hat{\phi}_{0}\right)^{2j-1}\, , \label{eq:c5s4e7}
\end{equation}%
which is the Laplace transform of (\ref{eq:c5s4e5}). The method
for determining $\hat{\phi}(s)$ is shown to be valid in this case.

(b) \textit{Comb structure with} $l=2a$. The secondary branch is
two-sites long, $A^{\prime}$ and $A^{\prime \prime}$. Similarly to
the previous case we can write the distribution for the time
probabilities as
\begin{equation}
\hat{\phi}(s)=\alpha \hat{\phi}_{0}\sum_{j=0}^{\infty }\left[\frac{%
(1-\alpha )}{2}\hat{\phi}_{0}^{2}\sum_{k=0}^{\infty }
\left(\frac{1}{2}\hat{\phi}_{0}^{2}\right) ^{k}\right]^{j}
=\frac{\alpha\hat{\phi}_{0}(2-\hat{\phi}_{0}^{2})} {2-(2-\alpha
)\hat{\phi}_{0}^{2}}\, . \label{eq:c5s4e7b}
\end{equation}

In this equation a new sum over the index $k$ appears, because the
walker can move away from the backbone twice. For each such
occurrence we must apply rule (i). We also assume that jumps to
the nearest neighbor occur with probability $1/2$ on the linear
teeth.

(c) \textit{Comb structure with} $l\rightarrow\infty$. Each time
the walker moves away from the backbone, a new convolution factor
appears in $\hat{\phi}(s)$. For the case $l\rightarrow \infty$, we
have in principle infinitely many convolution factors in the
expression for $\hat{\phi}(s)$. Fortunately, we can simplify this
situation considerably. Assume that the walker is at the first
site on the secondary branch, point $A^{\prime}$ in Fig.
\ref{fig:comb}, and moves away from the backbone. Let
$\eta_{A^{\prime}}$ be the probability distribution of returning
for the first time to the point $A^{\prime}$ after a time $t$. Now
suppose the same situation but for the initial point
$A^{\prime\prime}$. It is easy to see that as $l\rightarrow\infty$
the limit $\eta_{A^{\prime\prime}}\rightarrow\eta_{A^{\prime}}$
has to hold, and we can again use the rules (i) -- (iii) to
determine $\eta_{A^{\prime }}$. Doing so, we obtain the expression
\begin{equation}
\eta_{A^{\prime}}=\frac{1}{2}\hat{\phi}_{0}\sum_{j=0}^{\infty}
\left(\frac{1}{2}\hat{\phi}_{0}\eta_{A^{\prime \prime }}
\right)^{j}\, . \label{eq:c5s4e8}
\end{equation}%
This expression is equivalent to the form (\ref{eq:c5s4e6}) with
$\alpha=1/2$; on the secondary branch every jump to a nearest
neighbor occurs with probability $1/2$. Introducing the condition
$\eta_{A^{\prime\prime}}=\eta_{A^{\prime}}$, which is strictly
correct for $l=\infty$, and solving (\ref{eq:c5s4e8}), we find
\begin{equation}
\eta_{A^{\prime}}=
\frac{1-\sqrt{1-\hat{\phi}_{0}^{2}}}{\hat{\phi}_{0}}\, .
\label{eq:c5s4e9}
\end{equation}
With this result, the distribution $\hat{\phi}(s)$ is obtained
straightforwardly from the three rules (i) -- (iii),
\begin{equation}
\hat{\phi}(s)=\alpha \hat{\phi}_{0}\sum_{j=0}^{\infty}
\left[(1-\alpha )\hat{\phi}_{0}\eta _{A^{\prime }}\right]^{j}
=\frac{\alpha\hat{\phi}_{0}} {\alpha +\left( 1-\alpha \right)
\sqrt{1-\hat{\phi}_{0}^{2}}}\, . \label{eq:c5s4e10}
\end{equation}

In finding the waiting time pdf associated with the comb
structures, it is assumed that we are only interested in the
dynamical behavior along the backbone
and the rest of the structure is considered
as secondary.

\begin{figure}
\centerline{\includegraphics[scale=1]{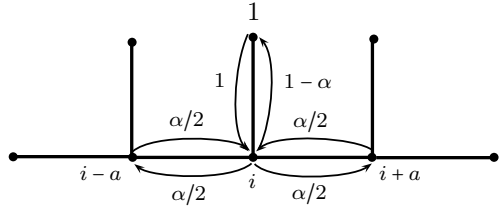}}
\caption{Comb with a single node along the tooth. In the picture
are written the probabilities of jumping between nodes of length
$a$.} \label{fig:combs}
\end{figure}

The second method consists in finding the master equation for the
random walk moving along the backbone. This master equation has to
have incorporated the movement along the teeth. Consider first the
simplest case of a one sided comb with a single node as shown in
Figure  \ref{fig:combs}

Let $\alpha$ be the probability of moving along the backbone, when the
walker is at a node of the backbone and $1-\alpha$ is the probability
of jumping to the teeth, if the walker is at the backbone. So, if the
movement along the backbone is considered isotropic, the probability
of jumping to the right or to the left is $\alpha/2$. The master
equation for the probability of finding a walker located at the node
$i$ of the backbone at time $t$ is (see Figure \ref{fig:combs})
\begin{equation}
P(i,t)=\frac{\alpha}{2}P(i+a,t-\tau)+
\frac{\alpha}{2}P(i-a,t-\tau)+P(1,t-\tau)\, ,\label{eq:me1}
\end{equation}
where we have taken into account that the walker waits a constant
time $\tau$ at every node between consecutive jumps. The master equation
for the node 1 of the tooth reads
\begin{equation}
P(1,t)=(1-\alpha)P(i,t-\tau)\label{eq:me2}
\end{equation}
that is coupled to (\ref{eq:me1}). Since time is a discrete variable
it is convenient to transform the time coordinate to new variable
$z$ through the transformation
\[
P(i,z)=\sum_{t=0}^{\infty}z^{t}P(i,t)\, .
\]
Multiplying (\ref{eq:me1}) and (\ref{eq:me2}) by $z^{t}$ and
summing from $t=0$ to infinity we have, respectively
\begin{equation}
P(i,z)=\frac{\alpha b}{2}P(i+a,z)+\frac{\alpha
b}{2}P(i-a,z)+bP(1,z)\, ,\label{eq:me3}
\end{equation}

\begin{equation}
P(1,z)=b(1-\alpha)P(i,z)\, ,\label{eq:me4}
\end{equation}
where $b=z^{\tau}$. Inserting (\ref{eq:me4}) into (\ref{eq:me3})
and rearranging terms we get the following master equation for the
movement along the backbone
\begin{equation}
\frac{1}{2}P(i+a,z)+\frac{1}{2}P(i-a,z)=
P(i,z)\frac{1-(1-\alpha)b^{2}}{\alpha b}\, .\label{eq:me5}
\end{equation}
It is convenient to stress that (\ref{eq:me5}) is actually the master
equation for a walker moving on a comblike constructed by repeating
the element depicted in Figure \ref{fig:combs}. Finally, we can write
Eq. (\ref{eq:me5}) in the Laplace space for time by taking into account
that $z=e^{-s}$, so that $b=e^{-s\tau}.$ Hence, (\ref{eq:me5})
becomes
\begin{equation}
\frac{1}{2}\hat{P}(i+a,s)+\frac{1}{2}\hat{P}(i-a,s)=
\hat{P}(i,s)\frac{1-(1-\alpha)e^{-2s\tau}}{\alpha e^{-s\tau}}\, .
\label{eq:me6}
\end{equation}

\begin{figure}
\centerline{\includegraphics[scale=0.8]{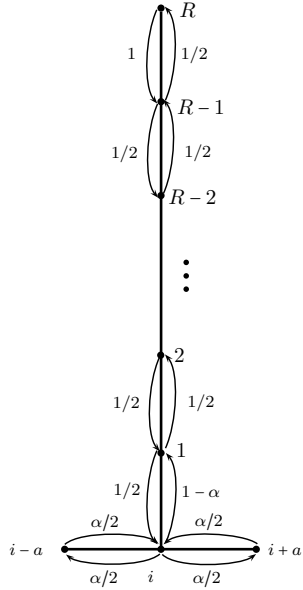}}
\caption{Comb with $R$ nodes along the teeth. In the picture are
written the probabilities of jumping between nodes of length $a$.}
\label{fig:4}
\end{figure}

The example consists in generalizing the above structure to a
tooth with $R$ nodes or length $l=aR$. Then, we consider a one
sided comb with whose basic element is given in Figure
\ref{fig:4}. The master equation for the probability of finding a
walker at the node $i$ of the backbone at time $t$ is, at the $z$
space

\begin{equation}
P(i,z)=\frac{\alpha b}{2}P(i+a,z)+\frac{\alpha b}{2}
P(i-a,z)+\frac{b}{2}P(1,z)\, .\label{eq:me7}
\end{equation}
At the node $R$ the master equation is
\begin{equation}
P(R,z)=\frac{b}{2}P(R-1,z)\, ,\label{eq:me8}
\end{equation}
and at the nodes $R-1$ and 1 they are
\begin{equation}
P(R-1,z)=\frac{b}{2}P(R-2,z)+bP(R,z)\, ,\label{eq:me9}
\end{equation}
\begin{equation}
P(1,z)=b(1-\alpha)P(i,z)+\frac{b}{2}P(2,z)\, .\label{eq:me9b}
\end{equation}
At a generic node $j$ of the tooth we get
\begin{equation}
P(j,z)=\frac{b}{2}P(j-1,z)+\frac{b}{2}P(j+1,z)\, ,\label{eq:me10}
\end{equation}
where $j=2,\ldots,R-2$. We may solve (\ref{eq:me10}) by proposing the
solution $P(j,z)=A\lambda^{j}.$ Inserting this solution we obtain
the characteristic equation $\lambda^{2}-2\lambda/b+1=0$. Hence,
\begin{equation}
P(j,z)=A_{1}\lambda_{+}^{j}+A_{2}\lambda_{-}^{j}\, ,\label{eq:sol}
\end{equation}
where $A_{1,2}$ are constant to be determined and
\[
\lambda_{\pm}=\frac{1}{b}\left(1\pm\sqrt{1-b^{2}}\right)\, .
\]
On setting $j=1$ and $j=2$ into Eq. (\ref{eq:sol}), we get the
system of algebraic equations for the constants $A_{1,2}$
\[
A_{1}\lambda_{+}+A_{2}\lambda_{-}=P(1,z)\, ,
\]
\begin{equation}
A_{1}\lambda_{+}^{2}+A_{2}\lambda_{-}^{2}=P(2,z)\, .
\label{eq:sis1}
\end{equation}
On the other hand, by inserting (\ref{eq:sol}) into (\ref{eq:me8})
and (\ref{eq:me9}) we get the equation relating $A_{1}$ and $A_{2}$
\begin{equation}
A_{1}\left[\lambda_{+}^{R-2} \left(1-\frac{3b^{2}}{4}\right)-
\frac{b}{2}\left(1-\frac{b^{2}}{2}\right)\lambda_{+}^{R-3}\right]+
A_{2}\left[\lambda_{-}^{R-2}\left(1-\frac{3b^{2}}{4}\right)-
\frac{b}{2}\left(1-\frac{b^{2}}{2}\right)\lambda_{-}^{R-3}\right]=0\,
. \label{eq:sis2}
\end{equation}
By combining (\ref{eq:sis2}) and (\ref{eq:sis1}) we can express
$P(2,z)$ in terms of $P(1,z)$ in the form $P(2,z)=h(b)P(1,z)$,
where
\begin{equation}
h(b)=\frac{1}{b}\left[2-\frac{b^{2}(\lambda_{+}^{R}+
\lambda_{-}^{R})}{\lambda_{+}^{R}+\lambda_{-}^{R}+
(\lambda_{-}^{R}-\lambda_{+}^{R})\sqrt{1-b^{2}}}\right]\, .
\label{eq:h}
\end{equation}
Now, we are in position to get the master equation for the motion
along the backbone by substituting $P(2,z)=h(b)P(1,z)$ into (\ref{eq:me9b})
and the result into (\ref{eq:me7}). The final result is
\begin{equation}
\frac{1}{2}P(i+a,z)+\frac{1}{2}P(i-a,z)= \frac{1}{\alpha
b}\left[1-\frac{b^{2}(1-\alpha)}{2-bh(b)}\right]P(i,z)\, .
\label{eq:me11}
\end{equation}
We show now how to reduce the effect of the teeth on a waiting
time distribution for the motion of a walker along the backbone.
To this end we make use of the CTRW, and in particular, of the
generalized master equation for finding a walker at point $x$ at
time $t$ when it moves a 1D space
\begin{equation}
\frac{\partial P}{\partial t}=\int_{0}^{t}K(t-t')\left[\int
P(x-x',t-t')w(x')dx'-P(x,t')\right]dt'\, . \label{eq:me12}
\end{equation}
Considering the jump lengths distribution given in
(\ref{eq:c5s4e1}), for jumps between the consecutive nodes of a 1D
lattice with spacing $a$ and transforming into the Laplace space
in time, Eq. (\ref{eq:me12}) becomes
\begin{equation}
\frac{1}{2}\hat{P}(i+a,s)+\frac{1}{2}\hat{P}(i-a,s)=
\hat{P}(i,s)\left[\frac{s+\hat{K}(s)}{\hat{K}(s)}\right]\, ,
\label{eq:me13}
\end{equation}
where the memory kernel $K(t)$ is related to the waiting time
distribution $\phi(t)$ through their Laplace transforms
\begin{equation}
\hat{K}(s)=\frac{s\hat{\phi}(s)}{1-\hat{\phi}(s)}\, .\label{mK}
\end{equation}
Therefore, Eq. (\ref{eq:me13}) reduces to
\begin{equation}
\frac{1}{2}\hat{P}(i+a,s)+\frac{1}{2}\hat{P}(i-a,s)=
\hat{P}(i,s)\frac{1}{\hat{\phi}(s)}\, .\label{eq:me14}
\end{equation}
The waiting time distribution is obtained by comparing Eqs.
(\ref{eq:me14}) and (\ref{eq:me11}), to get
\begin{equation}
\hat{\phi}(s)=\frac{\alpha b}{1-(1-\alpha)
\left[1+\frac{\lambda_{-}^{R}-\lambda_{+}^{R}}{\lambda_{-}^{R}+
\lambda_{+}^{R}}\sqrt{1-b^{2}}\right]}\, , \label{eq:wt}
\end{equation}
where $b=e^{-s\tau}$. If the one sided comb has a teeth with one
node, $R=1$, thus from Eq. (\ref{eq:wt}), the waiting time
distribution is
\[
\hat{\phi}(s)=\frac{\alpha b}{1-(1-\alpha)b^{2}}\, ,
\]
which coincides with the result obtained in Eq. (\ref{eq:c5s4e6}).
When $R=2$, then
\[
\hat{\phi}(s)=\frac{\alpha b(2-b^{2})}{2-(2-\alpha)b^{2}}\, ,
\]
which is the same result obtained in (\ref{eq:c5s4e7b}).
Finally, let us consider the infinite length of the teeth
$R\rightarrow\infty$, or $l\gg a$. In this case, Eq. (\ref{eq:wt})
reduces to
\begin{equation}
\hat{\phi}(s)=\frac{\alpha b}{\alpha+(1-\alpha)\sqrt{1-b^{2}}}\, ,
\label{eq:ad}
\end{equation}
that is equal to (\ref{eq:c5s4e10}). In the limit of large times,
$s\rightarrow0$, the waiting time distribution in Eq.
(\ref{eq:ad}) reads
$\hat{\phi}(s)\simeq1-(1-\alpha)\sqrt{2s\tau}/\alpha$, which
predicts anomalous diffusion along the backbone. So that, the
teeth have to have an infinite length to predict anomalous
diffusion along the whole comb {at the asymptotically large
times}.

\section{Comb-like models mimic spiny dendrites}
\begin{figure}
\centerline{\includegraphics[scale=0.8]{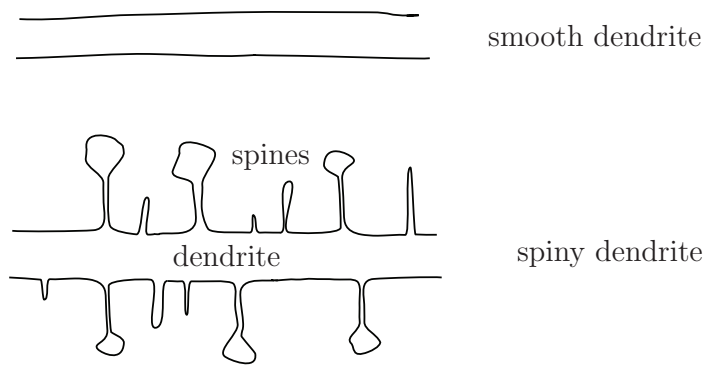}}
\caption{Draw of a smooth dendrite and a spiny dendrites}
\label{fig:5}
\end{figure}

As shown in previous studies, the geometric nature of spiny
dendrites plays essential role in kinetics
\cite{SaWi06,SaWi11,ZeHo11,bkh,bbz,kubota}. The
real distribution of spines along the dendrite, their size and
shapes are completely random \cite{nim}, and inside spines, not
only the spine necks but the spine itself acts as a transport
barrier {\cite{SaWi11,kubota,mic}}. Therefore a reasonable
assumption is a consideration of anomalous diffusion along both
the spines and dendrite. So, we propose models based on a
comb-like structure that mimics a spiny dendrite, where the
backbone is the dendrite and teeth (lateral branches) are the
spines, see Fig. \ref{fig:5} (We distinguish between a smooth and
a spiny dendrite). In this case dynamics inside teeth corresponds
to spines, while the backbone describes diffusion along dendrites.
Note that the comb model is an analogue of a 1D medium where
fractional diffusion has been observed and explained in the
framework of the CTRW \cite{WeHa86,ArBa91,Metzler20001,luba} and making use of macroscopic
descriptions {\cite{zabu}} .

Before embarking for the CTRW consideration in the framework of the comb
model, let us explain how anomalous diffusion in the comb model relates to the
CaMKII transport along the spiny dendrite, and how geometry of the latter
relates to the anomalous transport. As admitted above, the spine cavities
behave as traps for the contaminant transport. As follows from a general
consideration of a Markov process inside a finite region, the pdf of lifetimes inside the cavity with a finite volume and arbitrary form decays exponentially with time $t$ (see, for example,
{\cite{Re01}}) $\varphi (t) = \frac{1}{\tau} \exp (- \frac{t}{\tau}$). Here
$\tau$ is a survival time (mean life time), defined by the minimum eigenvalue
of the Laplace operator and determined by geometry of the cavity. For example,
in Refs. {\cite{bbz,kubota}}, for spines with a head of volume $V$ and the
cylindrical spine neck of the length $L$ and radius $a$, the mean life time is
$\tau = LV / \pi a^2 D = L^2 / D$, where $D$ is diffusivity of the spine.
Therefore, the mean probability to find a particle inside the spine after time
$t$, (i.e, the survival probability inside the cavity from 0 to $t$) averaged over all possible realizations of $\tau$ is given by the integral
\begin{equation}
  \label{gimp1} \Psi (t) = \int_t^{\infty} \int_0^{\infty} \varphi (t' / \tau) f
  (\tau) d \tau dt'
\end{equation}
where $f (\tau)$ is a distribution function of the survival times $\tau$
(recall that size and shape of spines are random {\cite{nim}}).  Finally, the waiting time pdf can be easily calculated from Eq.
(\ref{gimp1}), as follows
\begin{equation}
  \label{gimp11} \phi (t)=-\partial _t\Psi (t) =  \int_0^{\infty} \varphi (t / \tau) f
  (\tau) d \tau  .
\end{equation}
In the
simplest case, when the distribution is the exponential $f (\tau) = (1 /
\tau_0) \exp (- \tau / \tau_0$), one obtains from Eq. (\ref{gimp11}) that the
general kinetics is not Markovian and the waiting time pdf is a
stretched exponential for large times
\[ \phi (t) =\frac{1}{\tau _0}\int _0^{\infty}\frac{e^{-t/\tau}}{\tau}e^{-\tau/\tau _0}d\tau=\frac{2}{\tau _0}K_0\left( 2\sqrt{t/\tau _0}\right) \sim \left( \frac{t}{\tau_0} \right)^{- \frac{1}{4}} \exp (-
   \sqrt{t / \tau_0}) \hspace{0.25em}, t / \tau_0 \gg 1 \hspace{0.25em}
   . \]
The situation is more interesting, when the distribution of the survival times
is the power law $f (\tau) \sim 1 / \tau^{1 + \alpha}, ~ (0 < \alpha < 1)$. In
this case the waiting time pdf is the power law as well $\phi (t) \sim 1 /
t^{1 + \alpha}$ that leads to subdiffusion motion along the dendrite. This
result follows from the CTRW theory, since all underlying micro-processes are
independent Markovian ones with the same distributions
{\cite{Metzler20001}}.

Now we explain the physical reason of the possible power law distribution
$\phi (t)$. At this point we paraphrase some arguments from Ref. \cite{bAHa05}
with the corresponding adaptation to the present analysis. Let us consider the
escape from a spine cavity from a potential point of view, where geometrical
parameters of the cavity can be related to a potential $U$. For example, for
the simplest case, mentioned above, it is $U = VL / \pi a^2$, which ``keeps''
a particle inside the cavity, while $D \tau_0$ plays a role of the kinetic
energy, or the ``Boltzmann temperature''. Therefore, escape probability from
the spine cavity/well is described by the Boltzmann factor $\exp (- U /
D \tau_0)$. This value is proportional to the inverse waiting, or survival
time
\begin{equation}
  \label{gimp2} t \sim \exp \left( \frac{U}{D \tau_0} \right) .
\end{equation}
As admitted above, potential $U$ is random and distributed by the Poisson
distribution $P (U) = U_0^{- 1} \exp (- U / U_0)$, where $U_0$ is an averaged
geometrical spine characteristic. The probability to find the waiting time in
the interval $(t, t + dt)$ is equal to the probability to find the trapping
potential in the interval $(U, U + dU)$, namely $\phi (t) dt = P (U) dU$.
Therefore, from Eq. (\ref{gimp2}) one obtains
\begin{equation}
  \label{gimp3} \phi (t) \sim \frac{1}{t^{1 + \gamma}} \hspace{0.25em} .
\end{equation}
Here $\gamma = \frac{D \tau_0}{U_0} \in (0, 1)$ establishes a relation between
geometry of the dendrite spines and subdiffusion observed in
{\cite{SaWi06,SaWi11}} and support application of the comb
model, which is a convenient implement for analytical exploration of anomalous
transport in spiny dendrites in the framework of the CTRW consideration.

\subsection{Anomalous diffusion in spines}
Geometry of the comb structure makes it
possible to describe anomalous diffusion in  spiny dendrites
structure in the framework of the comb model.

Usually, anomalous diffusion on the comb is described by the $2D$
distribution function $P=P(x,y,t)$, and a special behavior is that
the displacement in the $x$--direction is possible only along the
structure backbone ($x$-axis at $y=0$). Therefore, diffusion in
the $x$-direction is highly inhomogeneous. Namely, the diffusion
coefficient is $D_{xx}=D_x\delta(y)$, while the diffusion
coefficient in the $y$--direction (along teeth) is a constant
$D_{yy}=D_y$. Due to this geometrical construction, the flux of
particles along the dendrite is
\begin{equation}\label{Res_1a}
J_{x}=-D_{x}\delta (y)\frac{\partial P}{\partial x}
\end{equation}
and the flux along the finger describes the anomalous trapping
process that occurs inside the spine
\begin{equation}\label{Res_1b}
J_{y}=\left. -D_{y}\frac{\partial ^{1-\gamma }}{\partial
t^{1-\gamma }}\right\vert_{RL}\frac{\partial P}{\partial y}
\end{equation}
where $P(x,y,t)$ is the density of particles and
\begin{equation}\label{der_RL}
\left.\frac{\partial^{1-\gamma}}{\partial t^{1-\gamma
}}\right\vert_{RL}f(t)=\frac{\partial}{\partial t}I_t^{\gamma}f(t)
\end{equation}
is the Riemann-Liouville fractional derivative, where the
fractional integration $I_t^{\gamma}$ is defined by means of the
Laplace transform
\begin{equation}\label{Res_1c}
\hat{\mathcal{L}}\left[I_t^{\gamma}f(t)\right]=
s^{-\gamma}\hat{f}(s)\, .
\end{equation}
So, inside the spine, the transport process is
anomalous and $\left\langle y^2(t)\right\rangle \sim t^{\gamma }$,
where $\gamma\in (0,1)$. Making use of the continuity equation for
the total number of particles
\begin{equation}\label{Res_1d}
 \frac{\partial P}{\partial t}+\mbox{div}\mathbf{J}=0\, ,
\end{equation}
where $\mathbf{J} = (J_{x},J_{y})$, one has the following evolution
equation for transport along the spiny dendrite
\begin{equation}\label{eb}
\frac{\partial P}{\partial t}-D_{x}\delta (y)\frac{\partial
^{2}P}{\partial x^{2}}-D_{y}\frac{\partial ^{1-\gamma }}{\partial
t^{1-\gamma }}\Big\vert_{RL} \frac{\partial ^{2}P}{\partial
y^{2}}=0.
\end{equation}
The Riemann-Liouville fractional derivative in Eq. (\ref{eb}) is
not convenient for the Laplace transform. To ensure feasibility of
the Laplace transform, which is a strong machinery for treating
fractional equations, one reformulates the problem in a form
suitable for the Laplace transform application.

To shed light on this situation, let us consider a comb in the
$3D$ {\cite{bAHa05}}. This model is described by the distribution
function $P_1(x,y,z,t)$ with evolution equation given by the equation
\begin{equation} \label{3d_comb1}
\frac{\partial P_1}{\partial t}-D_{x}\delta (y)\delta(z)
\frac{\partial ^{2}P_1}{\partial x^{2}}-
D_{y}\delta(z)\frac{\partial ^{2}P_1}{\partial y^{2}}-
\frac{\partial ^{2}P_1}{\partial z^{2}}=0\, .
\end{equation}%
It should be stressed that $z$ coordinate (do not confuse with the
$z$ variable introduced in the previous section) is a
supplementary, virtue variable, introduced to described fractional
motion in spines by means of the Markovian process. Thus the true
distribution is $P(x,y,t)=\int_{-\infty}^{\infty} P_1(x,y,z,t)dz$
with corresponding evolution equation
\begin{equation} \label{3d_comb2}
\frac{\partial P}{\partial t}-D_{x}\delta (y) \frac{\partial
^{2}P_1(z=0)}{\partial x^{2}}- D_{y}\frac{\partial
^{2}P_1(z=0)}{\partial y^{2}}=0\, .
\end{equation}
A relation between $P(x,y,t)$ and $P_1(x,y,z=0,t)$ can be
expressed through their Laplace transforms
\begin{equation}\label{3d_comb3}
\hat{P}_1(x,y,z=0,s)=\frac{\sqrt{s}}{2}\hat{P}(x,y,s)\, ,
\end{equation}
where $\hat{P}(x,y,s)=\hat{\mathcal{L}}[P(x,y,t)]$ and
$\hat{P}_1(x,y,z,s)=\hat{\mathcal{L}}[P_1(x,y,z,t)]$.

Equation (\ref{3d_comb3}) establishes a relationship between
the distributions $P_1(x,y,z=0,t)$ and $P(x,y,t)$ in the Laplace
space. Both distributions are related through the expression
\[ P \left( x, y, t \right) = \int_{- \infty}^{\infty} P_1 \left( x, y, z, t
   \right) d z\, . \]
If we transform the above equation by Fourier-Laplace we get
\begin{equation}
\hat{P} \left( k_x, k_y, s \right) = \hat{P}_1 \left( k_x, k_y, k_z =
   0, s \right)\, . \label{A1}
\end{equation}
Then, Eq. (\ref{3d_comb3}) is nothing but a relation between
$\hat{P}_1 \left( k_x,k_y, k_z = 0, s \right)$ and
$\hat{P}_1\left( k_x, k_y, z = 0, s \right)$. To find $\hat{P}_1
\left( k_x, k_y, k_z, s \right)$ we transform Eq. (\ref{3d_comb1})
by Fourier-Laplace and after collecting terms we find
\begin{equation}
 \hat{P}_1 \left( k_x, k_y, k_z, s \right) = \frac{1 - D_x k_x^2 P_1
   \left( k_x, y = 0, z = 0, s \right) - D_y k_y^2 P_1 \left( k_x, k_y, z = 0,
   s \right)}{s + k_z^2}\, , \label{A2}
\end{equation}
where the initial condition has been assumed $P_1 \left( x, y, z, t = 0
\right) = \delta \left( x \right) \delta \left( y \right) \delta \left( z
\right)$ for simplicity.
Setting $k_z = 0$ one gets
\begin{equation}
 \hat{P}_1 \left( k_x, k_y, k_z = 0, s \right) = \frac{1 - D_x k_x^2 P_1
   \left( k_x, y = 0, z = 0, s \right) - D_y k_y^2 P_1 \left( k_x, k_y, z = 0,
   s \right)}{s}\, .
  \label{A3}
\end{equation}
Inverting Eq. (\ref{A2}) by Fourier over $k_z$ we obtain
\[ \hat{P}_1 \left( k_x, k_y, z, s \right) = \frac{1 - D_x k_x^2 P_1 \left(
   k_x, y = 0, z = 0, s \right) - D_y k_y^2 P_1 \left( k_x, k_y, z = 0, s
   \right)}{2 \sqrt{s}} e^{- \sqrt{s} \left| z \right|}\, . \]
Then setting $z = 0$, one obtains
\begin{equation}
 \hat{P}_1 \left( k_x, k_y, z = 0, s \right) = \frac{1 - D_x k_x^2 P_1
   \left( k_x, y = 0, z = 0, s \right) - D_y k_y^2 P_1 \left( k_x, k_y, z = 0,
   s \right)}{2 \sqrt{s}}\, .
   \label{A4}
\end{equation}
Combining Eqs. (\ref{A3}) and (\ref{A4}), one has
\[ \hat{P}_1 \left( k_x, k_y, z = 0, s \right) = \frac{\sqrt{s}}{2}
   \hat{P}_1 \left( k_x, k_y, k_z = 0, s \right)\, , \]
then the Fourier inversion over $k_x$ and $k_y$ yields Eq.
(\ref{3d_comb3}). Finally, performing the Laplace transform of Eq.
(\ref{3d_comb2}) one obtains
\begin{equation}
s\hat{P}(x,y,s)-D_{x}\delta (y) \frac{\partial
^{2}\hat{P}_1(x,y,z=0,s)}{\partial x^{2}}- D_{y}\frac{\partial
^{2}\hat{P}_1(x,y,z=0,s)}{\partial y^{2}}=P(x,y,t=0)\, ,
\end{equation}
and substituting relation (\ref{3d_comb3}), dividing by $\sqrt{s}$
and then performing the Laplace inversion, one obtains the comb
model with the fractional time derivative
\begin{equation} \label{3d_comb4}
\frac{\partial^{\frac{1}{2}}P}{\partial t^{\frac{1}{2}}}-
D_{x}\delta (y)\frac{\partial ^{2}P}{\partial x^{2}}-D_{y}
\frac{\partial ^{2}P}{\partial y^{2}}=0 \, ,
\end{equation}
where $2D_{x,y}\rightarrow D_{x,y}$ and the Caputo
derivative\footnote{To avoid any confusion between the
Riemann-Liouville and the Caputo fractional derivatives, the
former one stands in the text with an index RL: $\frac{\partial
^{\alpha}}{\partial t^{\alpha}}\vert_{RL}$, while the latter
fractional derivative is not indexed $\frac{\partial
^{\alpha}}{\partial t^{\alpha}}$. Note, that it is also convenient
to use Eq. (\ref{LCaputo}) as a definition of the Caputo
fractional derivative.} $\frac{\partial^{\gamma}}{\partial
t^{\gamma}}$ can be defined by the Laplace transform for
$\gamma\in(0,1)$ \cite{Mainardi19961461}
\begin{equation}\label{LCaputo}
\hat{\mathcal{L}}\left[\frac{\partial^{\gamma}f}{\partial
t^{\gamma}}\right]= s^{\gamma}\hat{f}(s)-s^{\gamma-1}f(t=0)\, .
\end{equation}
The fractional transport takes place in both the
dendrite $x$ direction and the spines $y$ coordinate. To make
fractional diffusion in dendrite normal, we add the fractional
integration $I_t^{1-\gamma}$ by means of the Laplace transform
(\ref{Res_1c}), as well
$\hat{\mathcal{L}}\left[I_t^{1-\gamma}f(t)\right]=
s^{\gamma-1}\hat{f}(s)$. This yields Eq. (\ref{3d_comb4}), after
generalization $\frac{1}{2}\rightarrow\gamma\in(0,1)$,
\begin{equation} \label{3d_comb5}
\frac{\partial^{\gamma}P}{\partial t^{\gamma}}-
D_{x}\delta(y)I_t^{1-\gamma} \frac{\partial ^{2}P}{\partial
x^{2}}-D_{y} \frac{\partial ^{2}P}{\partial y^{2}}=0  .
\end{equation}
Performing the Fourier-Laplace transform in (\ref{3d_comb5}) we
get
\begin{equation}
P(k_{x},k_{y},s)=\frac{P(k_{x},k_{y},t=0)-D_{x}k_{x}^{2}P(k_{x},y=0,s)}{%
s+D_{y}k_{y}^{2}s^{1-\gamma }}\, ,  \label{tfl}
\end{equation}%
where the Fourier-Laplace image of the distribution function is
defined by its arguments
$\hat{\mathcal{L}}\hat{\mathcal{F}}_x\hat{\mathcal{F}}_y[P(x,y,t)]=P(k_x,k_y,s)$.
If $P(x,y,t=0)=\delta (x)\delta (y)$, inversion by Fourier over
$y$ gives
\begin{equation}\label{Res_2a}
P(k_{x},y,s)=\frac{1-D_{x}k_{x}^{2}P(k_{x},y=0,s)}{s^{(2-\gamma )/2}\sqrt{%
D_{y}}}\exp \left( -\left\vert y\right\vert s^{\gamma /2}/\sqrt{D_{y}}%
\right) .
\end{equation}
Taking $y=0$ the above equation provides%
\begin{equation}\label{Res_2b}
P(k_{x},y=0,s)=\frac{1}{s^{(2-\gamma )/2}
\sqrt{D_{y}}+D_{x}k_{x}^{2}}\, ,
\end{equation}
which yields after inserting into  Eq. (\ref{tfl})
\begin{equation}
P(k_{x},k_{y},s)=\frac{1}{s+D_{y}k_{y}^{2}s^{1-\gamma }}\left(
1-\frac{D_{x}k_{x}^{2}}{s^{(2-\gamma )/2}
\sqrt{D_{y}}+D_{x}k_{x}^{2}}\right)\, . \label{solf}
\end{equation}%
We can calculate the density of particles at a given point $x$ of
the dendrite at time $t$, namely $P(x,t)$, by integrating over $y$
in the Fourier space
\begin{equation}
P(k_{x},s)=P(k_{x},k_{y}=0,s)=\frac{s^{-\gamma /2}\sqrt{D_{y}}%
}{s^{(2-\gamma )/2}\sqrt{D_{y}}+D_{x}k_{x}^{2}}\, ,  \label{Pks}
\end{equation}%
then
\begin{equation} \label{msd_a}
\left\langle x^{2}(s)\right\rangle =-\left. \frac{\partial
^{2}}{\partial
k_{x}^{2}} P(k_{x},s)\right\vert _{k_{x}=0}=\frac{2D_{x}}{\sqrt{D_{y}}}\frac{1%
}{s^{2-\frac{\gamma }{2}}}
\end{equation}
so that
\begin{equation} \label{moment2}
\left\langle x^{2}(t)\right\rangle =\frac{2D_{x}}{\sqrt{D_{y}}}t^{1-\frac{%
\gamma }{2}}\, .
\end{equation}
Equation (\ref{moment2}) predicts subdiffusion along the spiny
dendrite that is in agreement with the experimental results
reported in \cite{SaWi06}. It should be noted that this
result is counterintuitive. Indeed, subdiffusion in spines, or
fingers should lead to the slower subdiffusion in dendrites, or
backbone with the transport exponent less than in usual comb,
since these two processes are strongly correlated. But this
correlation is broken due to the fractional integration
$I_t^{1-\gamma}$ in Eq. (\ref{3d_comb5}). On the other hand, if we
invert (\ref{Pks}) by Fourier-Laplace we obtain the fractional
diffusion equation for $P(x,t)$
\[
\frac{\partial ^{1-\frac{\gamma }{2}}P}{\partial t^{1-\frac{\gamma }{2}}}=%
\frac{D_{x}}{\sqrt{D_{y}}}\frac{\partial ^{2}P}{\partial x^{2}}\,
, \]%
which is equivalent to the generalized Master equation
(\ref{eq:me12}), in the diffusion limit:
\begin{equation} \label{ME_a}
\frac{\partial P}{\partial t}=\int_{0}^{t}K(t-t^{\prime
})\frac{\partial ^{2}P(x,t^{\prime })}{\partial x^{2}}dt^{\prime
}\, ,
\end{equation}%
if the Laplace transform of the memory kernel is given by $
\hat{K}(s)=\frac{D_{x}}{\sqrt{D_{y}}}s^{\gamma /2}$, which corresponds
to the waiting time pdf in the Laplace space given by
\begin{equation} \label{pdf}
\hat{\phi} (s)=\frac{1}{1+\frac{\sqrt{D_{y}}}{D_{x}}s^{1-\frac{\gamma
}{2}}}
\end{equation}
that is $ \phi (t)\sim t^{-2+\frac{\gamma }{2}}$ as
$t\rightarrow \infty$.  Let us employ the notation for a
dynamical exponent $d_w$ used in
\cite{SaWi06,SaWi11}. If $d_w=4/(2-\gamma)$ then
the MSD grows as $t^{2/d_w}$. On the other hand, it has been found
in experiments that  $d_w$ increases with the density of spines
$\rho _s$ and the simulations prove that $d_w$ grows linearly with
$\rho _s$. Indeed, the experimental data admits almost any growing
dependence of $d_w$ with $\rho _s$ due to the high variance of the
data (see Fig 5.D in \cite{SaWi11}). Equation
(\ref{moment2}) also establishes a phenomenological relation
between the second moment and $\rho _s$. When the density spines
is zero then $\gamma=0$, $d_w=2$ and normal diffusion takes place.
If the spine density $\rho _s$ increases, the anomalous exponent
of the pdf (\ref{pdf}) $1-\gamma/2=2/d_w$ must decrease (i.e., the
transport is more subdiffusive due to the increase of $\rho _s$)
so that $d_w$ has to increase as well. So, our model predicts
qualitatively that $d_w$ increases with $\rho _s$, in agreement
with the experimental results in \cite{SaWi11}.

\subsection{ L\'evy walks on fractal comb}
In this section we consider a fractal comb model \cite{iomin2011} to
take into account the inhomogeneity of the spines distribution.
Here, we consider the comb model for a phenomenological explanation  of an
experimental situation, where we introduce a control parameter
which establishes a relation between diffusion along dendrites and
the density of spines. Suggesting a more sophisticated relation
between the dynamical exponent and the spine density, we can
reasonably suppose that the fractal dimension, due to the box
counting of the spine necks, is not integer: it is embedded in the
$1D$ space, thus the spine fractal dimension is $\nu\in(0,1)$.
According the fractal geometry (roughly speaking), the most
convenient parameter is the fractal dimension of the spine volume
(mass) $\mu_{\rm spine}(x)\equiv\mu(x) \sim |x|^{\nu}$. Therefore,
following Nigmatulin's idea on a construction of a ``memory
kernel'' on a Cantor set in the Fourier space $|k|^{1-\nu}$
\cite{NIG92} (and further developing in Refs.
\cite{NIG98,Ren,bi2011}), this leads to a convolution integral
between the non-local density of spines and the probability
distribution function $P(x,y,t)$ that can be expressed by means of
the inverse Fourier transform \cite{iomin2011}
$\hat{\mathcal{F}}_x^{-1}\left[|k_x|^{1-\nu}P(k_x,y,t)\right]$.
Therefore, the starting mathematical point of the phenomenological
consideration is the fractal comb model
\begin{equation} \label{nu_comb10}
\frac{\partial^{\gamma}P}{\partial t^{\gamma}}-
D_{x}\delta(y)I_t^{1-\gamma} \frac{\partial ^{2}P}{\partial x^{2}}
-D_y\frac{\partial^2}{\partial y^2} \hat{\mathcal{F}}_{k_x}^{-1}
\left[|k_x|^{1-\nu}P(k_x,y,t)\right]=0\, .
\end{equation}
Performing the same analysis in the Fourier-Laplace space,
presented in previous section, then Eq. (\ref{Pks}) reads
\begin{equation} \label{nuPks}
P(k_{x},s)=P(k_{x},k_{y}=0,s)=
\frac{s^{-\gamma/2}\sqrt{D_{y}}}{s^{(2-\gamma )/2}
\sqrt{D_{y}}+D_{x}|k_{x}|^{\beta}}\, ,
\end{equation}
where $\beta=3/2+\nu/2$.

Contrary to the previous analysis expression (\ref{msd_a}) does
not work any more, since superlinear motion is involved in the
fractional kinetics. This leads to divergence of the second moment
due to the L\'evy flights. The latter are described by the
distribution $\sim 1/|x|^{1+\beta}$, which is separated from the
waiting time probability distribution $\phi(t)$. To overcome
this deficiency, we follow the analysis of the L\'evy walks
suggested in \cite{ZKB433,ZK1818}. We consider our
exact result in Eq. (\ref{nuPks}) as an approximation obtained
from the joint distribution of the waiting times and the L\'evy
walks. Therefore, a cutoff of the L\'evy flights is expected at
$|x|=t$. This means that a particle moves at a constant velocity
inside dendrites not all times, and this laminar motion is
interrupted by localization inside spines distributed in space by
the power law.

Performing the inverse Laplace transform, we obtain solution in
the form of the Mittag-Leffler function \cite{batmen}
\begin{equation}\label{Pkt}
P(k_x,t)= E_{1-\gamma/2}
\left(-D|k|^{\beta}t^{1-\gamma/2}\right)\, ,
\end{equation}
where $D=\frac{D_x}{\sqrt{D_y}}$. For the asymptotic behavior
$|k|\rightarrow 0$ the argument of the Mittag-Leffler function can
be small. Note that in the vicinity of the cutoff $|x|=t$ this
corresponds to the large $t$ ($|k|\sim\frac{1}{t}\ll 1$),  thus we
have \cite{batmen}
\begin{equation}\label{MLasympt}
E_{1-\gamma/2}
\left(-D|k|^{\beta}t^{1-\gamma/2}\right)\approx \exp\left(
-\frac{D|k|^{\beta}t^{1-\gamma/2}}{\Gamma(2-\gamma/2)}\right)\, .
\end{equation}
Therefore, the inverse Fourier transform yields
\begin{equation} \label{solPxt}
P(x,t)\approx A_{\gamma,\nu}
\frac{Dt^{1-\gamma/2}}{\Gamma(2-\gamma/2)|x|^{(5+\nu)/2}}\, ,
\end{equation}
where $A_{\gamma,\nu}$ is determined from the normalization
condition (The physical plausibility of estimations
(\ref{MLasympt}) and (\ref{solPxt}) also follows from the
plausible finite result of Eq. (\ref{solPxt}), which is the
normalized distribution $P(x,t)\sim 1/|x|^{(3+\nu+\gamma)/2}$,
where $|x|=t$). Now the second moment corresponds to integration
with the cutoff at $x=t$ that yields
\begin{equation} \label{msd_b}
\left\langle x^{2}(t)\right\rangle =
K_{\gamma,\nu}t^{\frac{3-\gamma-\nu}{2}}\, ,
\end{equation}
where
$K_{\gamma,\nu}=\frac{4A_{\gamma,\nu}D_{x}}{(1-\nu)\Gamma(2-\gamma/2)\sqrt{D_{y}}}$
is a generalized diffusion coefficient. Transition to the absence of
spines means first transition to normal diffusion in teeth with
$\gamma=1$ and then $\nu=0$ that yields
\begin{equation}\label{END}
\left\langle x^{2}(t)\right\rangle =K_{1,0}t\, .
\end{equation}

\subsection{Fractional reaction-diffusion along spiny dendrites }
Geometrically, spiny dendrites in the $3D$ space are completely
described by a comb structure in the  $2D$, where the spine
density on the cylinder surface is projected on the $1D$ axis (say
the $x$ axis): $\rho(x,r={\rm const},\theta)\rightarrow\rho(x)$.
Here $\rho(x,r=const,\theta)$ is the spine density, while $\rho(x)$
is the density of the comb teeth. In what follows, we consider
$\rho(x)=g={\rm const}$, which is, probably, the most realistic
case. Fractional diffusion inside the spines is described by
fractional diffusion inside the teeth. Therefore, one considers a
two-sided comb model as in Figure \ref{fig:1}, and the starting
mathematical point of the phenomenological consideration is the
Fokker-Planck equation obtained in \cite{MeIo13}.

It reads,
\begin{equation}\label{A11}
\frac{\partial P}{\partial t} -\delta(y)\frac{\partial^2
P}{\partial x^2} -g\frac{\partial^2P}{\partial y^2}=0\, .
\end{equation}
This equation is obtained by the re-scaling with relevant combinations of the
comb parameters $D_x$ and $D_y$, such that the dimensionless
time and coordinates are $D_x^3t/D_y^2\rightarrow t$
$D_x/D_y\rightarrow x$, $D_x/D_y\rightarrow y/\sqrt{g}$,
correspondingly \cite{ib2005}, and parameter $g$ can be considered
as a constant density of the fingers.

As admitted above a variety of interactions inside spines leads to
correlated noises in dendritic spines \cite{Zeng}. The strong
correlations of that leads to anomalous (subdiffusive) motion
inside the spines. Following a phenomenological description by the
CTRW, this subdiffusion is controlled by a waiting-time
pdf $\phi (t)$ decaying
according to the power law. Therefore, normal diffusion of the
contaminant density $P(x,y,t)$, for example activated CaMKII, in
spines is replaced by the anomalous transition term
\begin{equation}\label{A21}
g\frac{\partial^2 P}{\partial y^2}\Rightarrow
g\int_0^t K(t-t')\frac{\partial^2 P(t')}{\partial
y^2}dt'\, ,
\end{equation}
where $K(t)$ is again the time memory kernel of the generalized
master equation and is defined in (\ref{mK}). For subdiffusion,
$\phi (t)=\frac{1}{1+t^{1+\gamma}}$ with $0<\gamma<1$
that yields \cite{sokolov} $\hat{K}(s)=s^{1-\gamma}$.

One may recognize that Eq. (\ref{A21}) is a formal
expression for the anomalous transport with very complicated form
in the time domain, which in turn, is very inconvenient for
the analytical treatment. Therefore, the comb model may be
presented in the Laplace domain. Substituting Eq. (\ref{A21}) in
Eq. (\ref{A11}), then performing the Laplace transform and taking
into account Eq. (\ref{mK}), one obtains the comb model in the
Laplace domain
\begin{equation}\label{A41}
s\hat{P}=\delta(y)\frac{\partial^2 \hat{P}}{\partial x^2}
-gs^{1-\alpha}\frac{\partial^2\hat{P}}{\partial y^2}+P_0\, .
\end{equation}
Here $P_0=P(x,y,t=0)$ is the initial condition. As admitted, the
kernel $\hat{K}(s)$ is problematic for the Laplace
inversion, since it leads to the appearance of the initial
condition. To overcome this obstacle, one multiplies Eq.
(\ref{A41}) by $s^{\alpha-1}$ and then perform the Laplace
inversion that yields
\begin{equation}\label{A5}
\int_0^t(t-t')^{-\gamma}\left[\frac{\partial
P(x,y,t')}{\partial t'}- \delta(y)\frac{\partial
^{2}P(x,y,t')}{\partial
x^{2}}\right]dt'=
g\frac{\partial^2P(x,y,t)}{\partial y^2}\, .
\end{equation}
Amending this equation by reaction term, one arrives at the
integro-differential equation:
\begin{equation}\label{nu_comb0}
\int_0^t(t-t')^{-\gamma}\left[\frac{\partial
P(x,y,t')}{\partial t'}- \delta(y)\frac{\partial
^{2}P(x,y,t')}{\partial
x^{2}}\right]dt'=
g\frac{\partial^2P(x,y,t)}{\partial
y^2}+g\hat{C}\left[P(x,y,t)\right]\, ,
\end{equation}
which describes the $2D$ inhomogeneous reaction diffusion in the
dispersive medium. Here $\hat{C}[P(x,y,t)]\equiv\hat{C}(P)$ is
a reaction kinetic term. In particular, to model reaction kinetics
inside dendrites, it can be considered either linear
$\hat{C}(P)=CP$, or logistic
$\hat{C}(P)=CP(x,y,t)[1-P(x,y,t)]$ \cite{b2}. Integration with
the power law kernel $t^{-\gamma}$ ensures anomalous diffusion in
both the dendrite and spines.

In what follows we use convenient notations of fractional
integro-differentiation given in Eq. (\ref{LCaputo}) and the text below.
Owing
to this notation, from Eq. (\ref{nu_comb0}) the equation for $P(x,y,t)$ reads
\begin{equation} \label{nu_comb1}
\frac{\partial^{\gamma}P}{\partial t^{\gamma}}-
\delta(y)I_t^{1-\gamma} \frac{\partial ^{2}P}{\partial x^{2}}
-g\frac{\partial^2 P}{\partial y^2} = g\hat{C}(P)\, .
\end{equation}

To give a first and brief insight on the problem of the front
propagation, let us consider the linear reaction and $\gamma=1$.
In this case, one obtains a ``simple'' solution \cite{iom2012} for
the travelling wave along the $x$ axis (inside dendrites).
Introducing the total probability distribution function
$P_1(x,t)=\int dy P(x,y,t)$, one obtains
\begin{equation}\label{sim_sol}
P_1(x,t)=\frac{\sqrt{2g^{1/2}}}{\pi\sqrt{t^{1/2}}}
\exp\left[\frac{x^2}{2\sqrt{gt}}-Cgt\right]\, .
\end{equation}
This yields the coordinates of the front $x \sim t^{3/4}$ that
spreads with the decaying velocity $v\sim t^{-1/4}$. This solution
illustrates the asymptotic failure of the reaction-transport front
propagation due to subdiffusion inside spiny dendrites.

\section{Front propagation in combs}
Recently, a mechanism of translocation wave of CaMKII  has been
suggested \cite{pcb}. As shown, activated CaMKII contaminant
travels along dendrites with additional translocation inside
spines. Process of activation (the conversion of primed CaMKII to
its active state) corresponds to the irreversible reaction that,
in absence of spines, is described by the Fisher-Kolmogorov-Petrovskii-Piskunov (FKPP) equation (It relates also to the logistic kinetic term)
\cite{pcb}. Therefore, in the framework of the suggested above
scheme of the dispersive subdiffusive comb (\ref{nu_comb1}),
nonlinear reaction at subdiffusion in dendrites takes place along
the $x$-axis bound, while subdiffusion in fingers describes the
translocation inside spines. Therefore, reaction-transport
equation (\ref{nu_comb1}) now reads
\begin{equation} \label{fkppc1}
\frac{\partial^{\gamma}P}{\partial t^{\gamma}} -
\delta(y)I_t^{1-\gamma} \left[D\frac{\partial ^{2}P}{\partial
x^{2}}+ \hat{C}(P)\right]  =g\frac{\partial^2}{\partial y^2} P\, .
\end{equation}
Here $D$ describes the diffusivity inside dendrites, while
$\hat{C}(P)=CP(1-P)$ is the nonlinear reaction term.  Again,
integrating over $y$ to obtain the kinetic equation for the total
distribution $P_1(x,t)$, we have
\begin{equation}\label{fkppc1a}
\frac{\partial^{\gamma}P_1}{\partial t^{\gamma}} - I_t^{1-\gamma}
\left[D\frac{\partial ^{2}P_0}{\partial x^{2}}+
\hat{C}(P_0)\right] =0\, .
\end{equation}
For the brevity, we denoted $P_0=P(x,y=0,t)$.
Consider the fractional comb model (\ref{fkppc1})
without reaction,
\begin{equation} \label{B2}
\frac{\partial^{\gamma}P}{\partial t^{\gamma}} -
\delta(y)I_t^{1-\gamma} D\frac{\partial ^{2}P}{\partial x^{2}}
=g\frac{\partial^2}{\partial y^2} P\, .
\end{equation}
and perform the Laplace transform, thus one obtains
\begin{equation}\label{B3}
s^{\gamma}\hat{P}-\delta(y)Ds^{\gamma-1}\frac{\partial^2
\hat{P}}{\partial x^2}=g\frac{\partial^2\hat{P}}{\partial y^2}
+s^{\gamma-1}\delta(x)\delta(y),
\end{equation}
where for the initial condition we take
$P(t=0)=\delta(x)\delta(y)$. Looking for the
solution in form
\begin{equation}\label{B3_2}
\hat{P}(x,y,s)=\exp[-\sqrt{s^{\gamma}/g}|y|]f(x,s)\, .
\end{equation}
one can see that
$\hat{P}_0=\hat{P}(x,y=0,s)=f(x,s)$  and integration
Eq. (\ref{B3_2}) over $y$ yields
\begin{equation} \label{B1}
\hat{P}_0(x,s)=\sqrt{\frac{s^{\gamma}}{4g}}\hat{P}_1(x,s).
\end{equation}

Substituting (\ref{B1}) into (\ref{fkppc1a}) in the Laplace space, one gets
\begin{equation}\label{fkppc3}
s\hat{P}_1-P_1(t=0)-\frac{Ds^{\frac{\gamma}{2}}}{2\sqrt{g}}\frac{\partial
^{2}\hat{P}_1}{\partial x^{2}}
-\frac{Cs^{\frac{\gamma}{2}}}{2\sqrt{g}}\hat{P}_1=-C\hat{\mathcal{L}}[P_0^2]
\, .
\end{equation}
Multiplying this equation by $e^{st}$ and using identity
$e^{st}s^{\alpha}f(s)=\frac{\partial}{\partial t}
e^{st}s^{\alpha-1}f(s)$, we integrate with the corresponding
contour to obtain the inverse Laplace transform. This yields
\begin{equation}\label{fkppc4}
\frac{\partial P_1}{\partial t}-\frac{1}{2\sqrt{g}} \frac{\partial
}{\partial t}I_t^{1-\frac{\gamma}{2}}\left[D\frac{\partial
^{2}P_1}{\partial x^{2}}+CP_1\right]=-\frac{C}{4g}\left[
\frac{\partial }{\partial t}I_t^{1-\frac{\gamma}{2}}P_1\right]^2\,
.
\end{equation}
The nonlinear term is obtained by the following chain of
transformations
\begin{equation}
C[P_0^2]=C\left[\hat{\mathcal{L}}^{-1}\hat{P}_0\right]^2=
\frac{C}{4g}\left[\hat{\mathcal{L}}^{-1}s^{\frac{\gamma}{2}}\hat{P}_1\right]^2
=\frac{C}{4g}\left[\frac{\partial }{\partial
t}I_t^{1-\frac{\gamma}{2}}P_1\right]^2\, .
\end{equation}
Note that a specific property of these transformations is an
irreversibility with respect to the Laplace transform, since, as
well known, the Laplace transform of the Riemann-Liouville
fractional derivative involves the (quasi) initial value terms like
$P_1(t=0)=\delta(x)$ \cite{Metzler20001}.


To evaluate the overall velocity of the asymptotic front, let us
introduce a small parameter, say $\varepsilon$, at the derivatives
with respect to the time and space \cite{Fr}. To this end we
re-scale $x\rightarrow x/\varepsilon$ and $t\rightarrow
t/\varepsilon$, and $P_1(x,t)\rightarrow
P_1^{\varepsilon}(x,t)=P_1\left(\frac{x}{\varepsilon},
\frac{t}{\varepsilon}\right)$. Therefore, one looks for the
asymptotic solution in a form of the Green approximation,
\begin{equation}\label{WKB}
P_1^{\varepsilon}(x,t)=
\exp\left[-\frac{G^{\varepsilon}(x,t)}{\varepsilon}\right]\, .
\end{equation}
The main strategy of implication of this construction is the limit
$\varepsilon\rightarrow 0$, one has
$\exp\left[-\frac{G^{\varepsilon}(x,t)}{\varepsilon}\right]=0$,
except the condition when $G^{\varepsilon}(x,t)=0$. This equation
determines the position of the reaction spreading front, see Eq. (\ref{sim_sol}).
Moreover, we consider the limit $G(x,t)=\lim_{\varepsilon\to
0}G^{\varepsilon}(x,t)$ as the principal Hamiltonian function
\cite{Fr} that makes it possible to apply the Hamiltonian
approach for calculation of the propagation front velocity. In
this case partial derivatives of $G(x,t)$ with respect to time and
coordinate have physical senses of the Hamiltonian and momentum:
\begin{equation}\label{Action}
\frac{\partial G(x,t)}{\partial t}=-H~~~~ \mbox{and}~~~~
\frac{\partial G(x,t)}{\partial x}=p\, .
\end{equation}

Now the method of the hyperbolic scaling, explained above, can be
applied. Therefore, we have the ansatz (\ref{WKB}) for the
probability distribution function inside dendrites. Inserting
expression (\ref{WKB}) in Eq. (\ref{fkppc4}), one considers
fractional integrations in time. Let us start from the last term
in Eq. (\ref{fkppc4}), which is the reaction term. We rewrite it
in the following convenient form
\begin{equation}\label{react_a}
\frac{\varepsilon}{\Gamma(1-\frac{\gamma}{2})}\frac{\partial}{\partial
t} \int_0^{\frac{t}{\varepsilon}}dt'(t^{\prime})^{-\gamma/2}
\exp[-G^{\varepsilon}(t-\varepsilon t',x-\varepsilon
x')/\varepsilon]\, .
\end{equation}
Then performing expansion $$G^{\varepsilon}(t-\varepsilon
t',x-\varepsilon x')\approx G^{\varepsilon}(x,t)-\varepsilon
\frac{\partial G^{\varepsilon}(x,t)}{\partial t}t'-\varepsilon
\frac{\partial G^{\varepsilon}(x,t)}{\partial x}x'\, ,$$ and
substituting this in Eq. (\ref{react_a}), one obtains
\begin{equation}\label{react_b}
\frac{1}{\Gamma(1-\frac{\gamma}{2})}\left[-\frac{\partial
G^{\varepsilon}(x,t)}{\partial t}\right]
\exp\left[-\frac{G^{\varepsilon}(x,t)}{\varepsilon}\right]
\int_0^{\frac{t}{\varepsilon}}(t^{\prime})^{-\gamma/2}
\exp\left[\frac{\partial G^{\varepsilon}(x,t)}{\partial
t}t'+\frac{\partial G^{\varepsilon}(x,t)}{\partial x}x'\right]dt'
\, .
\end{equation}
It should be noted that we neglect differentiation of the upper
limit of the integral, since this term is of the order of
$O(\varepsilon^{1+\gamma/2})\sim o(\varepsilon)$ that vanishes in
the limit $\varepsilon\rightarrow 0$. The same procedure of
expansion is performed for the diffusion term in Eq.
(\ref{fkppc4}) that yields
\begin{equation}\label{diff_a}
\frac{\varepsilon^3}{\Gamma(1-\frac{\gamma}{2})}
\frac{\partial^3}{\partial t\partial x^2}
\exp\left[-\frac{G^{\varepsilon}(x,t)}{\varepsilon}\right]
\int_0^{\frac{t}{\varepsilon}}(t')^{-\frac{\gamma}{2}}
\exp\left[\frac{\partial G^{\varepsilon}(x,t)}{\partial
t}t'+\frac{\partial G^{\varepsilon}(x,t)}{\partial x}x'\right]dt'
\, .
\end{equation}
Differentiating in the limit $\varepsilon\rightarrow 0$ and taking
into account that the Hamiltonian $H$ and the momentum $p$ in Eq.
(\ref{Action}) are independent of $x$ and $t$ explicitly (that
leads to the absence of mixed derivatives), one obtains the
Laplace transform of the subdiffusive kernel
$t^{-\frac{\gamma}{2}}$. After these procedures in  Eqs.
(\ref{react_b}) and (\ref{diff_a}), the kinetic equation
(\ref{fkppc4}) becomes a kind of Hamilton-Jacobi equation that
establishes a relation between the Hamiltonian and the momentum
\begin{equation}\label{fkppc5}
H=\left[\frac{Dp^2+C}{2\sqrt{g}}\right]^{\frac{2}{2-\gamma}}\, ,
\end{equation}
and the action is
$G(x,t)=\int_0^t[p(s)\dot{x}(s)-H(p(s),x(s))]ds$. The rate $v$ at
which the front moves is determined at the condition $G(x,t)=0$.
Together with the Hamilton equations, this yields
\begin{equation}\label{frontV}
v=\dot{x}=\frac{\partial H}{\partial p}\, ,~~~v=\frac{H}{p}\, . %
\end{equation}
Note that the first equation in (\ref{frontV}) reflects the
dispersion condition, while the second one is a result of the
asymptotically free particle dynamics, when the action is
$G(x,t)=px-Ht$. Taking into account $x=vt$, one obtains Eq.
(\ref{frontV}) (see also details of this discussion \textit{e.g.}
in Refs. \cite{CFM,ana}). Combination of these two equations can be
replaced by
\begin{equation}\label{minmin}
v=\min _{H>0}\frac{H}{p(H)}=\min _{p>0}\frac{H(p)}{p}\, .
\end{equation}
We also have from the front velocity conditions  (\ref{frontV})
$\frac{\partial}{\partial p}\ln H=1/p$ that, eventually,  yields
from Eq. (\ref{minmin})
\begin{equation}\label{fkppc6}
v=\left[\left(\frac{4}{g}\right)^{\frac{2}{2-\gamma}}\frac{D}{2-\gamma}
\left(\frac{C}{2+\gamma}\right)^{\frac{2+\gamma}{2-\gamma}}
\right]^{\frac{1}{2}} \, .
\end{equation}
To proceed, we first, admit that the limiting case of this result
with $\gamma=0$ corresponds to the CaMKII propagation along the
dendrite only (i.e., there are no teeth).
Therefore, Eq. (\ref{fkppc6}) after rescaling
$D/\sqrt{g}\rightarrow D$ and $C/\sqrt{g}\rightarrow C$ recovers
the FKPP scheme for $\gamma=0$ that yields $v=\sqrt{DC}$.

It should be admitted the absence of the failure of the activation
front propagation. It has a simple explanation due to the absence
of a reaction ``sink'' term  $-hP$ in Eq. (\ref{fkppc1}) by
neglecting the possibility of spines to accumulate a large amount
of ${\rm Ca}^{2+}$ \cite{segal1,segal2}, where $h$ is a
translocation/accumulation rate \cite{pcb}. Introducing this term
in Eq. (\ref{fkppc1}), our anticipation is that the hyperbolic
scaling for this new equation yields a solution similar to Eq.
(\ref{minmin}) with $H=0$ that corresponds to the failure of the
front propagation. Moreover, this asymptotic solution for
$P_1(x,t)$ always takes place, as one of possible solutions.

Inserting the sink term in Eq. (\ref{fkppc1}), one obtains
\begin{equation} \label{fkppc7}
\frac{\partial^{\gamma}P}{\partial t^{\gamma}}-
\delta(y)I_t^{1-\gamma} \left[D\frac{\partial ^{2}P}{\partial
x^{2}}+ CP(1-P)\right]-g\frac{\partial^2P}{\partial y^2}-ghP=0\, .
\end{equation}
Repeating the same procedures of the Laplace transform and integration over $y$
with definition
$\hat{P}_1=\int_{-\infty}^{\infty}\hat{P}(x,y,s)dy$, and using
the substitute $$P_1(x,s)=2\sqrt{g/s^{\gamma}}P(x,y=0,s)\, , $$
one obtains
\begin{equation}\label{fkppc8}
s\hat{P}_1-\delta(x)=\frac{D\sqrt{s^{\gamma}}}{2\sqrt{g}}
\frac{\partial\hat{P}_1}{\partial
x^2}+\frac{C\sqrt{s^\gamma}}{2\sqrt{g}}\hat{P}_1
-hgs^{1-\gamma}\hat{P}_1\, .
\end{equation}
Here we neglect the nonlinear term, since, as follows from the
above analysis, in the further hyperbolic scaling approximation,
this term does not contribute to the Hamilton-Jacobi equation, and
we also know how to handle it. Again multiplying this equation by
$e^{st}$ and using the same identity
$e^{st}s^{\alpha}f(s)=\frac{\partial}{\partial t}
e^{st}s^{\alpha-1}f(s)$, as above, we obtain the inverse Laplace
transform. Thus, Eq. (\ref{fkppc8}) reads
\begin{equation}\label{fkppc9}
\frac{\partial P_1}{\partial t} =
\frac{D}{2\sqrt{g}}\frac{\partial}{\partial t}
I_t^{1-\frac{\gamma}{2}}\left[ \frac{\partial P_1}{\partial x^2}
+\frac{C}{D}P_1\right]-hg\frac{\partial}{\partial
t}I_t^{\gamma}P_1\, .
\end{equation}
Application of the hyperbolic scaling with the asymptotic solution
(\ref{WKB}) yields
\begin{equation}\label{fkppc10}
2\sqrt{g}H=\left[Dp^2H^{\frac{\gamma}{2}}+CH^{\frac{\gamma}{2}}
-2hg^{\frac{3}{2}}H^{1-\gamma}\right]\, .
\end{equation}
Let us consider a specific case $\gamma=2/3$ that yields
\begin{equation}\label{fkppc11}
H=\left[\frac{Dp^2+C-2hg^{\frac{3}{2}}}{2\sqrt{g}}\right]^{\frac{3}{2}} .
\end{equation}
For $C>2hg^{\frac{3}{2}}$ there is no failure and the front
asymptotically propagates with a constant velocity. For $C\leq
2hg^{\frac{3}{2}}$ the only solution is $H=0$ and yields $v=0$.
So, $2hg^{\frac{3}{2}}$ is the minimum reaction rate necessary to
sustain propagation along the spiny dendrite due to the presence
of translocation. Analogously, $ (C/2h)^{2/3}$ can also be viewed
as the minimum value for the density of spines necessary to have
propagation failure. Both results are in agreement with the
results obtained from very different models based on the cable
model \cite{cb}.

In general case, one compares the interplay between the activation
$CH^{\frac{\gamma}{2}}$ and the translocation
$-2hg^{\frac{3}{2}}H^{1-\gamma}$ terms in Eq. (\ref{fkppc10}) in
the limit $H\rightarrow 0$. For $\gamma\in [\frac{2}{3},1)$, the
translocation term is dominant and leads to the solution with
$H=0$ and the failure of the front propagation, correspondingly.
When $0<\gamma<\frac{2}{3}$, the activation in dendrites can be
dominant. This situation is more complicated, and the
activation-translocation front can propagate with an
asymptotically finite velocity.

Finally, let us consider the linear counterpart of Eq.
(\ref{fkppc1}) with the linear reaction term $\hat{C}(P)=CP$. This
analysis will be useful to uncover the behavior of the tail of the
total distribution and check if the front accelerates or travel
with constant velocity. Rewrite the equation for the total
distribution $P_1(x,t)$. As follows from the fractional
differentiation of Eq. (\ref{fkppc4}), this equation reads (see
also \cite{MeIo13})
\begin{equation}\label{estim1}
\frac{\partial^{1-\frac{\gamma}{2}}P_1}{\partial
t^{1-\frac{\gamma}{2}}}=\frac{D}{2\sqrt{g}}\left[\frac{\partial^2}
{\partial x^2}+\frac{C}{D}\right]P_1\, ,
\end{equation}
with the initial condition $P_1(x,t=0)=\delta(x)$. After the Fourier
transform $\hat{\mathcal{F}}[P_1(x,t)]=\bar{P}_1(k,t)$, one
obtains the solution in the form of the Mittag-Leffler function
\begin{equation}\label{estim2}
\bar{P}_1(k,t)=E_{1-\frac{\gamma}{2}}
\left[A(k)t^{1-\frac{\gamma}{2}}\right]\, ,
\end{equation}
where $A(k)=(C-Dk^2)/2\sqrt{g}$. At the asymptotic condition, when
$x\, , t\gg 1$, we have $C\gg Dk^2$ that yields asymptotic
behavior of the Mittag-Leffler function as a growing exponent (for
the large positive argument) \cite{batmen}
\begin{equation}\label{estim3}
\bar{P}_1(k,t)\approx \exp\left[
\left(\frac{C}{2\sqrt{g}}-\frac{D}{2\sqrt{g}}k^2\right)^{\frac{2}{2-\gamma}}t\right]
\approx\exp\left[\left(\frac{C}{2\sqrt{g}}\right)^{\frac{2}{2-\gamma}}
\left(1-\frac{2}{2-\gamma}\frac{Dk^2}{C}\right)t\right]\, .
\end{equation}
After the Fourier inversion, one obtains
\begin{equation}\label{estim4}
P_1(x,t)=\exp\left[\left(\frac{C}{2\sqrt{g}}\right)^{\frac{2}{2-\gamma}}t
-\frac{(2-\gamma)x^2(2\sqrt{g})^{\frac{2}{2-\gamma}}}
{8DC^{\frac{\gamma}{2-\gamma}}t}\right]
\end{equation}
that, finally, yields the nonzero and constant overall velocity of
the reaction front propagation. Note that for normal diffusion,
$\gamma=0$, one arrives at the Fisher velocity
$v=\sqrt{DC/g}\rightarrow\sqrt{DC}$, see limiting case $\gamma=0$
in Eq. (\ref{fkppc6}).

\section{Conclusion}
In this chapter we show that a comb is a convenient model for
analytical exploration of anomalous transport and front
propagation phenomena along spiny dendrites. We have studied the
properties of a random walk motion along the backbone in presence of
teeth. Teeth are lateral branches crossing the backbone and we
have shown here that their effect on the movement along the whole
structure can be reduced to a waiting time distribution at the
nodes of the backbone during the movement of the random walker.
Recent experiments and numerical simulations have predicted
anomalous diffusion along spiny dendrites.

We have shown here that this anomalous phenomenon can be explained in the framework of a comb model with infinitely long teeth. Moreover, due to the random distribution of spines along the parent dendrite and the presence of binding reaction inside spines one can present a physically reasonable justification of the power law of the waiting time distribution that leads to subdiffusion in both the teeth and the backbone.

 We have shown how to predict anomalous diffusion in
spines by constructing a fractional-diffusion equation. By using the
CTRW formalism we have computed the mean square displacement for
the transport along the whole comb. We presented an illustration of how
to take into account the inhomogeneous distribution of spines
along the dendrite by using a fractal comb. We have also
constructed the corresponding fractional-diffusion equations and
computed the mean square displacement. From the other hand, the constructed toy models are simple enough, like the comb model that makes it possible to suggest and understand a variety of
reaction-transport schemes, including anomalous transport, by
applying a strong machinery of fractional calculus and hyperbolic
scaling for asymptotic methods. This approach allows to suggest
an analytical description of reaction-transport scenarios in spiny
dendrites, where we consider both a linear reaction in spines, see
Eqs. (\ref{nu_comb0}) and (\ref{nu_comb1}), and nonlinear reaction
along dendrites, considered in the framework of the FKPP scheme
\cite{Fisher1937}. To this end we suggest a fractional
subdiffusive comb model, where we apply a Hamilton-Jacobi approach
to estimate the overall velocity of the reaction front
propagation. We proposed an alternative approach of a recently
suggested mechanism of translocation wave of CaMKII \cite{pcb},
where activated CaMKII contaminant travels along dendrites with
additional translocation inside spines, and  process of activation
corresponds to the irreversible reaction described by the FKPP
equation (\ref{fkppc4}). One of the main effect, observed in the
framework of the considered model, is the failure of the front
propagation due to either the reaction inside spines, or
interaction of reaction with spines. In the first case the spines
are the source of reactions, while in the latter case the spines
are a source of damping, for example they act as a sink of an
activated contaminant (CaMKII). The situation is controlled by
three parameters CaMKII activation $C$, CaMKII translocation rate
$h$ and the fractional transport exponent $\gamma$. The latter
reflects the geometrical structure of the transport system: when
$0<\gamma<\frac{2}{3}$, the activation in dendrites can be
dominant, and the activation-translocation front can propagate
with an asymptotically nonzero and constant velocity. For $\gamma
=2/3$ we have found a criteria for the emergence of propagation
failure or for the sustain of the propagation in terms of the
reaction rate, the translocation rate and the spine's density.

It should be admitted, in conclusion, that physical arguments
suggested above, explain why anomalous transport, namely
subdiffusion, of either CaMKII or neutral particles is possible
and support implementation of the comb model. These arguments are
based on geometry of dendritic spines that determines an
expression for the transport exponent in Eq. (\ref{gimp3}). This
situation becomes more sophisticated in a case of the nonlinear
FKPP reaction. Indeed, as shown, the power law kernel of the
transition probability considered due to the geometry arguments is
insensitive to the nonlinear reaction. This consideration differs
completely from a mesoscopic non-Markovian approach, developed in
\cite{prl,MeFeHo10}, where spines-dendrite interaction and an
extension including reactions in spines have been described in
framework of variable residence time. This leads to essential
complication of the transition probability due to the nonlinear
reaction term \cite{fedotov2010a,MeFeHo10}.


\section{Appendix: Fractional integro--differentiation}\label{sec:app_A}

The consideration of a non-Markovian process in the framework of
kinetic equations leads to the study of the so-called fractional
Fokker-Planck equation, where both time and space processes are
not local \cite{Metzler20001}. In this case,  derivations are
substituted by integrations with the power law kernels. One
arrives at so-called fractional integro--differentiation.

A basic introduction to fractional calculus can be found, {\em
e.g.}, in Ref. \cite{podlubny}. Fractional integration of the
order of $\alpha$ is defined by the operator
\begin{equation}\label{A1s}  %
{}_aI_t^{\alpha}f(t)=\frac{1}{\Gamma(\alpha)}
\int_a^tf(\tau)(t-\tau)^{\alpha-1}d\tau, ~~(\alpha>0)\, ,
\end{equation} %
where $\Gamma(\alpha)$ is a gamma function. There is no constraint
on the limit $a$. In our consideration, $a=0$ since this is a
natural limit for the time. A fractional derivative is defined as
an inverse operator to ${}_aI_t^{\alpha}\equiv I_t^{\alpha}$ as
$\frac{d^{\alpha}}{dt^{\alpha}}=I_t^{-\alpha}=D_t^{\alpha}$;
correspondingly $ I_t^{\alpha}=\frac{d^{-\alpha}}{dt^{-\alpha}}
=D_t^{-\alpha}$. Its explicit form is convolution
\begin{equation}\label{A2s}  %
D_t^{\alpha}=\frac{1}{\Gamma(-\alpha)}\int_0^t
\frac{f(\tau)}{(t-\tau)^{\alpha+1}}d\tau \, .
\end{equation}    %
For arbitrary $\alpha>0$, this integral is, in general, divergent.
As a regularization of the divergent integral, the following two
alternative definitions for  $D_t^{\alpha} $ exist \cite{Mainardi19961461}
\begin{equation}\label{A3s} %
{}_{RL}D_{(0,t)}^{\alpha}f(t)\equiv D_{RL}^{\alpha}f(t)=
D^nI^{n-\alpha}f(t) 
\frac{1}{\Gamma(n-\alpha)}\frac{d^n}{dt^n}\int_0^t
\frac{f(\tau)d\tau}{(t-\tau)^{\alpha+1-n}} \, ,
\end{equation} %
\begin{equation}\label{A4s} %
D_C^{\alpha}f(t)=
I^{n-\alpha}D^nf(t)  
\frac{1}{\Gamma(n-\alpha)}\int_0^t
\frac{f^{(n)d\tau}(\tau)}{(t-\tau)^{\alpha+1-n}} \, ,
\end{equation}  %
where $ n-1<\alpha<n,~~n=1,2,\dots$. Eq. (\ref{A3s}) is the
Riemann--Liouville derivative, while Eq. (\ref{A4s}) is the
fractional derivative in the Caputo form \cite{Mainardi19961461}.
Performing integration by part in Eq. (\ref{A3s}) and then applying
Leibniz's rule for the derivative of an integral and repeating
this procedure $n$ times, we obtain
\begin{equation}\label{A5s} %
D_{RL}^{\alpha}f(t)=D_C^{\alpha}f(t)+\sum_{k=0}^{n-1}f^{(k)}(0^+)
\frac{t^{k-\alpha}}{\Gamma(k-\alpha+1)} \, .
\end{equation}   %
The Laplace transform can be obtained for Eq. (\ref{A4s}). If
$\hat{\mathcal{L}}[f(t)]=\tilde{f}(s)$, then
\begin{equation}\label{A6}  %
\hat{\mathcal{L}}\left[D_C^{\alpha}f(t)\right]=s^{\alpha}\tilde{f}(s)-
\sum_{k=0}^{n-1}f^{(k)}(0^+)s^{\alpha-1-k}\, .
\end{equation}   %
The following fractional derivatives are helpful for the present
analysis
\begin{equation} \label{A8}
D_{RL}^{\alpha}[1]=\frac{t^{-\alpha}}{\Gamma(1-\alpha)}\, , ~~
D_C^{\alpha}[1]=0\, .
\end{equation} %
We also note that
\begin{equation}\label{A9}
D_{RL}^{\alpha}t^{\beta}=\frac{t^{\beta-\alpha}\Gamma(\beta+1)}
{\Gamma(\beta+1-\alpha)}\, ,
\end{equation}  %
where $\beta>-1$ and $\alpha>0$.
The fractional derivative from an exponential function can be
simply calculated as well by virtue of the Mittag--Leffler
function (see {\em e.g.}, \cite{podlubny,batmen}):
\begin{equation}\label{A10}   %
E_{\gamma,\delta}(z)=\sum_{k=0}^{\infty}
\frac{z^k}{\Gamma(\gamma k+\delta)} \, .
\end{equation}   %
Therefore, we have the following expression
\begin{equation}\label{A11s}   %
D_{RL}^{\alpha}e^{\lambda t}=t^{-\alpha}E_{1,1-\alpha}(\lambda
t)\, .
\end{equation}

\putbib

\clearpage
\printindex

\end{document}